# Potential Relationship of Epistemic Games to Group Dynamics and Learning Orientations towards Physics Problem Solving

Andrew J. Mason and Charles A. Bertram

*University of Central Arkansas, Department of Physics and Astronomy,
201 S. Donaghey Avenue, Conway, AR 72035*

**Abstract:** Current investigations into pedagogical goals of introductory algebra-based physics students at the University of Central Arkansas, by learning orientation towards an in-class metacognitive group problem solving task, seek to determine possible relationships with attitudinal shifts and course performance. Students thus far have been untreated with known group-based learning pedagogies, so as to establish trends of common group habits, and ultimately to properly inform implementation of group-based pedagogies in reaction to these trends. However, students' group dynamics and learning orientations prove difficult to map to group-based measurements; an estimate of group learning orientation and preferred working group dynamic is here explored as a potential means of interpreting students' use of problem solving strategies. A means of "sampling" audiovisual data in a live classroom of several simultaneous groups is also presented as a way to estimate the frequency of chosen strategies to this end.

## I. INTRODUCTION

Recent studies in algebra-based introductory physics laboratory sections have focused on attitudinal trends towards a metacognitive group problem solving exercise. [1-2] Since this course is predominately taken by either life science or health science majors, their attitudes towards physics may be sub-optimal if they do not value the course as they might value courses from their respective major tracks. We focused on group problem solving as a means of observing in real-time whether their approach towards the exercise accurately reflects their attitudes towards physics, and also whether this would have an effect upon their course grades.

In order to determine students' approach toward solving physics problems in groups, we now choose to look at structures of knowledge and strategy as employed into physics problem solving, which Tuminaro and Redish define as "epistemic games," [3-4] and consider how students' choice of games may affect how fruitful the problem solution is. Tuminaro and Redish note [3] that some games are more "intellectually complex" than other games, in that students would be more likely to develop conceptual understanding and less likely to engage in superficial learning. The games are named by Tuminaro and Redish, and placed in approximate order of most to least intellectually complex, as follows: Mapping Meaning to Mathematics (Mean-Math), Mapping Mathematics to Meaning (Math-Mean), Physical Mechanism (PM), Pictorial Analysis (PA), Recursive Plug and Chug (PC), and Transliteration to Mathematics (TM).

Two variables may factor into how students thus organize and use their knowledge for problem solving strategies. First, previous research [1-2] found tentative trends for how individual students may differ by major in terms of attitudes towards physics, and also how they may differ by learning orientation. [5] The results showed that these orientations appeared to reflect general changes in attitude towards the course as well as towards aspects of physics problem solving. However, it is difficult to trace how different individuals' attitudinal shifts may be reflected in a group activity. Since learning orientations appear to reflect attitudinal shifts towards the course, an initial approach is therefore to instead examine learning orientations that appear to drive a group's progress on a problem, as a possible factor influencing each group's choice and frequency of chosen epistemic games.

Second, the interactions with which students actually collaborate in a group, i.e. group dynamics, may also influence the choice and frequency of epistemic games. Groups that are not properly collaborative may opt to choose problem solving approaches that may be less useful, but instead more convenient for group members who are not participating in a productive struggle.

### A. Current Focus of Research

In this paper we address the working dynamics of lab groups, in an attempt to more clearly define different basic group dynamics, as well as specify the apparent driving learning orientations among each group's members, and determine whether either variable has an effect on lab groups' choice of epistemic games. Students' framing and approach towards a problem may be affected adversely if they are not truly working collaboratively, or if their learning orientations cause them to not have the proper interest in the exercise. We present an initial analysis of data taken from a first-semester algebra-based introductory physics course over three different semesters, each semester containing two laboratory sections of up to 24 students each. We will quantify the choice of epistemic games as an approximate percentage of sampled audiovisual data, and examine the potential relationship of the frequency of games to learning orientations and to basic interactive behaviors observed within different groups.

## II. PROCEDURE

### A. Group-Based Problem Solving Exercise

Three sections of a regional four-year state university's introductory algebra-based physics course, from the Spring 2014 (S14), Spring 2015 (S15), and Fall 2015 (F15) semesters, were chosen for the study. The department changed textbooks for the course during this time; the S14 course had a more traditional textbook, [6] while the S15 and F15 courses used a PER-based textbook. [7] Otherwise, instruction was the same for all courses: three 50-minute lecture sessions and one 3-hour laboratory session per week, all of which were taught by the same instructor. Each course initially contained 48 students divided into two 24-student laboratory sections.

In the first part of each laboratory section per week (aside from exam weeks), students were coached to solve a context-rich problem [8] with their lab partners over the course of about 45-50 minutes, and during this time, identify their strengths and weaknesses in their problem solving approach using a rubric adapted from a previous study on self-diagnosis of mistakes on exams. [9] During this time, the instructor and a Learning Assistant (LA) [10] proctored the classroom, providing assistance with solving the problem as needed. At the end of the period, work stopped and students were shown the solution by the instructor, whereupon they could finish their self-diagnosis rubrics if needed.

### B. Data Collection

Students gave voluntary written consent to be recorded during a problem solving session toward the end of the semester. The problem solving exercise was carried out as previously described; meanwhile, the instructor, LA, and 1-2 volunteer assistants would record all groups using hand-held cameras in brief (1-2 minute) samplings. There were 40 total students recorded in the S14 semester in this way, as well as 34 students in the S15 semester and 39 students in the F15 semester.

Learning orientation was determined by collection of written survey responses about students' perceived usefulness of the lab groups' problem solving exercise.

Three orientation categories emerged. "Framework-oriented" (F) students were explicitly interested in learning all or part of a problem solving framework (e.g. comments about improving one's visualization of the problem situation or approach to a solution). "Performance-oriented" (P) students were more interested in how the exercise helped them perform well in other aspects of the course (e.g. how the exercise helped one study for exams or do well on homework). The third category consisted of responses that appeared unrelated to describing overall course goals (e.g. comments about liking the opportunity to work with lab partners), and we tentatively defined it as a "vaguely-oriented" (V) category. Students' orientations, thus defined, could potentially be useful in terms of interpreting how student attitudes towards physics were applied towards a team effort at problem solving.

Due to unavoidable constraints on available cameras and camera users, care was taken to "randomly sample" students working together in their preferred collaboration, so as to observe chosen epistemic games [3] used by each group, and offer validity to the percentage of time taken for each specific game. Note that footage time per lab group was not always approximately equal. Certain groups interacted in more diverse ways than did others, which corresponded to requiring more footage. In addition, certain students who asked for help would get prolonged exposure to recording. In order to address this, plans were made to normalize the data with respect to total recording time for each group.

## III. INITIAL ANALYSIS

### A. Established Group Behavior Types

One of the factors involved in determining length of recording was in terms of how genuinely interactive a given laboratory group was; to that end, audiovisual data was observed for evidence of different group behaviors. Figure 1 shows the four main types of observed interactions that students exhibited while working in groups. The first type of interaction mainly consisted of not interacting at all, i.e. when students would choose to work independently (hereafter referred to as "I"). Students typically did this with the intention of only discussing the problem when stuck or having arrived at a solution. The second type dealt with a semi-collaborative ("S") effort, in which two students worked together while the third remained passive or worked individually. The third type was a fully collaborative ("C") effort where all students interacted to discuss the problem and make headway together. The fourth type involved a dominant group member simply directing the other students how to do the problem ("D"). Students did not necessarily remain in any one form of interaction, but could evolve over time; however, groups typically used a predominant dynamic over the others. Note that students could be in either groups of two or three; in the case of two students, there would be no semi-collaborative situations.

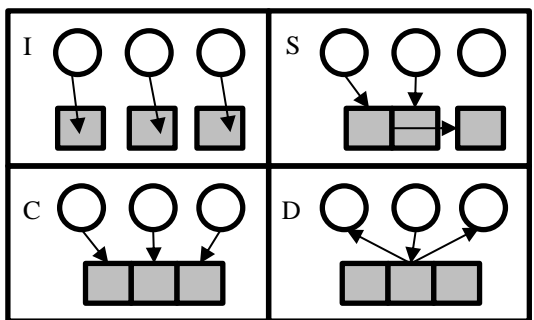

**FIG 1.** Four different types of observed laboratory group interaction. See text for descriptions.

### B. Analysis of Chosen Epistemic Games over Time

We normalized our sampled data for recorded epistemic games by counting the total number of games recorded for a given group, and then dividing by the total recording time in minutes. This ensured that there would not be a skewing of total number of epistemic games by the total recording time.

Figure 2 displays a chart of average games per minute taken for all groups for each semester (in order to account for any variance between courses), with standard uncertainty calculated for each mean value (~0.10 for all games together, ~0.01-0.08 for each individual game). Abbreviations in Section IIB are used for each game. All groups with at least 60 seconds of footage were included; the average amount of footage per group ranged about 400-600 seconds, with one section in the F15 semester getting 800-1000 seconds per group thanks to an additional available camera at that time.

Table I shows numerical values of games per minute for each bar in Figure 2. Statistical significance between semesters for each type of game was explored with a 2-tailed t-test; the main source of statistical significance occurred between the S14 semester and each of the other two semesters in the Transliteration to Mathematics game (p = 0.055 between S14 and S15; p = 0.004 between S14 and F15). Of interest is that the S14 class used a more traditional textbook [6] while the S15 and F15 classes used a PER-based textbook [7]; while it is encouraging that students make more use of a pedagogically informed textbook, it is one of the less conceptually deep games, [3] and contains a risk of superficial learning.

The F15 semester also had a significantly higher number of overall epistemic games per minute than did the S14 semester (p<0.001); and a borderline significantly higher number than did the S15 semester (p = 0.06). A one-way ANOVA between all three semesters over the average numerical values for each game, however, did not show an overall statistical significance between groups (F = 0.93, $F_{crit}$ = 3.68, p = 0.41).

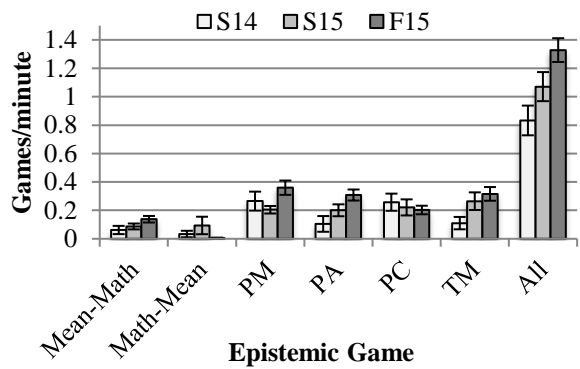

**FIG 2.** Number of games per minute for each type of epistemic game for the S14, S15, and F15 semesters.

TABLE I. Number of games per minute per group for each class.

| | Semester | S14 | S15 | F15 | All |
|---|---|---|---|---|---|
| | # Students | 40 | 35 | 39 | 114 |
| | # Groups | 13 | 14 | 15 | 42 |
| Games/min per | Mean-Math | 0.06 | 0.09 | 0.14 | 0.10 |
| | Math-Mean | 0.03 | 0.09 | 0.00 | 0.04 |
| | PM | 0.27 | 0.20 | 0.36 | 0.28 |
| | PA | 0.10 | 0.20 | 0.31 | 0.21 |
| | PC | 0.26 | 0.22 | 0.20 | 0.23 |
| | TM | 0.11 | 0.26 | 0.32 | 0.24 |
| | All | 0.83 | 1.07 | 1.33 | 1.08 |

### C. Relationship to Learning Orientations?

We next consider how the number of games per minute may be affected by preferred group dynamic or by the predominant driving learning orientation for a given laboratory group. We define the latter as either the orientation for the majority of members in each group, or failing a majority, the orientation of the student that seemed to lead the group's progress. This was applied to the F15 semester's students, of which there was relatively more footage recorded than the other two semesters, and who readily exhibited the behaviors in Fig. 1.

TABLE II. Number of games per minute, by predominant group dynamic and by driving learning orientation, for the F15 class. Number of total groups is in parentheses for each cell. "Other" refers to groups being driven by students with no recorded orientation.

|  |  | \multicolumn{5}{c}{Learning orientation driving group progress} |  |  |  |
|---|---|---|---|---|---|---|
|  |  | F | P | V | Other | All |
| Predominant group dynamic | I | - | 1.80 (1) | - | 1.09 (2) | 1.30 (3) |
|  | S | 1.59 (2) | 0.91 (1) | - | - | 1.34 (3) |
|  | C | 1.19 (2) | 1.38 (3) | 1.31 (3) | - | 1.29 (8) |
|  | D | 1.06 (1) | - | - | - | 1.06 (1) |
|  | All | 1.29 (5) | 1.37 (5) | 1.31 (3) | 1.09 (2) | 1.33 (15) |

Table II shows a distribution of games per minute among groups by driving orientation (columns) and by the most common group dynamic (rows) as described in Section IIIA. Examining Table II by group dynamic, it appears that the majority of examined groups were indeed able to collaborate as a team, either in full (C) or at least in part (S). There appears to be only a small effect on overall frequency of epistemic games per minute, either by learning orientation or by choice of group dynamic. Performance-oriented students appear to use slightly more games per minute, while group dynamics do not seem to cause much variance.

Most groups either had a performance- or framework-oriented dominance. The vague-dominated groups all seemed to favor collaboration; this reflects the strong tendency among vague-oriented students to discuss the process of working with partners rather than desired outcome. [2] Framework-dominated groups also tended to favor full or partial collaboration; the group with a dominant dynamic featured a framework-oriented student whose partners were not very strong in course performance.

## IV. DISCUSSION

We have demonstrated a potential means of "sampling" audiovisual data of a live laboratory classroom using brief recordings of each laboratory table, in terms of frequency of used epistemic games. This data may in turn be used to cross-reference the driving learning orientation for each group with the predominant group dynamic activity.

The number of games per minute may also provide a preliminary quantitative measure that can link the predominant group dynamic with the learning orientation that drives group progress. While total use of games does not appear to change much by driving learning orientation or predominant group dynamic, future analysis of individual games or of multiple semesters may show further differences.

Future planned analyses include an expansion of Table II to include data for additional semesters, in order to determine whether any significant differences may emerge with a larger sample size. We will also consider choice of individual games in the same vein as all total games were considered in Table II, e.g. whether certain orientations or group dynamics are more favorable for intellectually complex games.

## ACKNOWLEDGEMENTS

We thank M. Milan, B. Davanzo, and L. Ratz of the UCA SPS Chapter for their contributions in collecting data and discussions about data analysis. Funding was provided by the University of Central Arkansas Sponsored Programs Office and Department of Physics and Astronomy.